# Formulation and solutions of fractional continuously variable order mass spring damper systems controlled by viscoelastic and viscous-viscoelastic dampers

**S. Saha Ray**

Associate Professor;Department of Mathematics
National Institute of Technology
Rourkela-769008, India
Email: santanusaharay@yahoo.com

**S. Sahoo**

Research-Scholar; Department of Mathematics
National Institute of Technology
Rourkela-769008, India

**S. Das**

Scientist (H+); Reactor Control System Design Section, E & I Group
Bhabha Atomic Research Centre
Trombay-400085, Mumbai, India
Senior Research Professor, Dept. of Physics, Jadavpur University Kolkata
Adjunct Professor. DIAT-Pune
UGC Visiting Fellow. Dept of Appl. Mathematics; Univ. of Calcutta
Email: shantanu@barc.gov.in

## Abstract

The article presents theformulation and a new approach to find analytic solutions for fractional continuously variable order dynamic models viz. Fractional continuously variable order mass-spring damper systems. Here, we use the viscoelastic and viscous-viscoelastic dampers for describing the damping nature of the oscillating systems, where the order of fractional derivative varies continuously. Here, we handle the continuous changing nature of fractional order derivative for dynamic systems, which has not been studied yet. By successive iteration method, here we find the solution of fractional continuously variable order mass-spring damper systems, and then give a close form solution. We then present and discuss the solutions obtained in the cases with continuously variable order of damping for this oscillator with graphical plots.

## 1. Introduction

In past few years, the fractional order physical models [1, 2] have seen much attention by researchers due to dynamic behaviour and the viscoelastic behaviour of material [3]. Thus the fractional order model is remarkably used for describing the frequency distribution of the structural damping systems [4-7]. Several authors have modelled the dynamic system based on fractional calculus. Rossikhin and Shitikova [8] have done analysis on viscoelastic single-mass system by considering the damped vibration. Enelund and Josefson [9] used finite element method for analysis of fractionally damped viscoelastic material. The exact solution of fractional order of 1/2 was obtained by Elshehawey et al. [10]. The Green function approach for finding solution of dynamic system was studied by Agrawal [11], which followed by the Mittag-leffler function proposed by Miller [12]. By using fractional

Green function and Laplace transform, Hong et al. [13] has obtained the solution of single-degree freedom mass-spring system of order $0 < \alpha < 1$. The analytical solution of fractional systems mass-spring and spring-damper system formed by using Mittag-Leffler function was analysed by Gomez-Aguilar et al [14]. The fractional Maxwell model for viscous-damper model and its analytical solution was proposed by Makris and Constantinous [15], Choudhury et al. [16].

Other methods like Fourier transform [17-19] and Laplace transform [19-22] have been proposed by researchers to find the solution of fractional damper systems. Recently Saha Ray et al. [23] used the Adomian decomposition method to determine the analytical solution of dynamic system of order one-half and proclaim that the acquired solutions coincided with the solutions obtained through eigenvector expansion method given by Suarez and Shokooh [24]. Naber[25] used Caputo approach to study linear damping system. The generalization of linear oscillator to form the fractional oscillator has been studied by Stanislavsky [26]. The variable order structure is described in Laplace domain by Das [7]. Here in this article we extend the concept of continuously variable order structure of differential equation and get time domain solution.

The objectives of this article are first the mathematical formulation of fractional continuously variable order spring-mass damping systems and then analysing approximate analytical solution of fractional continuous-variable order models,in which damping are controlled by viscoelastic and viscous-viscoelastic dampers. Due to dynamic varying nature of fractional order derivative of damper material, it is very difficult to obtain the analytic solutions of the system. The solutions for fractional continuously variable dynamic models have been newly studied in this article. The linear damping natures of the systems have been taken here for modelling the problems. The changing property of the guide, on which the motion takes place, results in oscillation of the systems, which are modelled here by fractional continuously variable-order $q$. The obtained results have been plotted for showing the nature of oscillation, with continuously variable damping order.

The contents of the paper are organised as follows. Some mathematical aspects of fractional calculus including Riemann-Liouville approach and Mittag-Leffler function have been presented in Section 2. Section 3 presents the algorithm of successive iteration method. The dynamic fractional continuously-variable order mass-spring damping systems have been formulated in Section 4. The successful implementations of proposed successive iteration method for finding the analytical solutions of fractional dynamic systems have been discussed in Section 5. The numerical simulations for the results as obtained have been studied in Section 6. Section 7 concludes the paper.

## 2. Mathematical aspects of fractional calculus

### 2.1. Definition (Riemann-Liouville):

There are several definitions of fractional derivative [1, 2] that have been proposed in past. Here we review the most frequently used definitions of fractional integral viz. Riemann-Liouville integral, which is defined as follows:

$$J_t^{\alpha} f(t) = \frac{1}{\Gamma(\alpha)} \int_0^t (t-\tau)^{\alpha-1} f(\tau)\, d\tau \,,\; t > 0 \,,\; \alpha \in \Re^{+} \tag{2.1.1}$$

and the fractional derivative viz.Riemann-Liouville derivative of order $\alpha$ is defined as

$$D_t^{\alpha} f(t) = \frac{1}{\Gamma(m-\alpha)} \frac{d^m}{dt^m} \int_0^t (t-\tau)^{m-\alpha-1} f(\tau)\, d\tau \,,\; t > 0 \,,\; \alpha \in \Re^{+} \,, \tag{2.1.2}$$

where $m$ is positive integer, with $m - 1 \le \alpha < m$.

### 2.2. Definition (Mittag-Leffler function):

The two-parameter generalized Mittag-Leffler function [1, 2] defined by means of series expansion is as follows:

$$E_{\alpha,\beta}(z) = \sum_{k=0}^{\infty} \frac{z^k}{\Gamma(\alpha k + \beta)} \text{ for } \alpha > 0 \text{ and } \beta > 0 . \tag{2.2.1}$$

## 3. Basic principle of proposed successive iteration method

For defining the brief outline of proposed method, let us consider the differential equation in the following form

$$Lu + Ru = g \tag{3.1}$$

where $L$ and $R$ are invertible linear operator and remaining of the linear part respectively. The detailing of this method is described in Das [7], and Saha Ray et al [23], and the symbolic part described here-in this section will be clear in the solutions in subsequent sections.
The general solution of the eq. (3.1) can be written as

$$u = \sum_{n=0}^{\infty} u_n \tag{3.2}$$

where the complete solution of $Lu = g$ is $u_0$. By using the property of invertible linear operator, we can write the equivalent expression of eq. (3.2) as following form:

$$L^{-1}Lu = L^{-1}g - L^{-1}Ru \tag{3.3}$$

For initial value problem, we define the inverse linear operator $L^{-1}$ for $L = D_t^n$ that is *n*-fold derivative operator, and its inverse that is $L^{-1}$ will be *n*-fold integration operation from 0 to *t*. If we take $L = D_t^2$ then we have $L^{-1}Lu = u - u(0) - tu'(0)$. So eq. (3.3) reduces to

$$u = u(0) + tu'(0) + L^{-1}g - L^{-1}Ru \tag{3.4}$$

and for boundary value problem we have

$$u = A + Bt + L^{-1}g - L^{-1}Ru \tag{3.5}$$

where the integration constants $A$ and $B$ are determined from the given condition. Let $u_0 = u(0) + tu'(0) + L^{-1}g$.
So the general solution of eq. (3.1) becomes

$$u = u_0 - L^{-1}R\sum_{n=0}^{\infty} u_n \tag{3.6}$$

where

$$u_0 = \phi + L^{-1}g \tag{3.7}$$

and $\phi$ is the solution of $Lu = 0$, so that $L\phi = 0$.
So we have

$$u_{n+1} = -L^{-1}Ru_n, \qquad n \ge 0 \tag{3.8}$$

So by using eq. (3.8), we can find the $u_1, u_2, \ldots$ etc.

## 4. The problem formulation for mass-spring damper system

Damping is defined as restraining of vibratory or oscillatory motion; that means it reduces, restricts and prevents the oscillation of an oscillatory system. When the damping force is viscoelastic, it has both viscous and elastic characteristic to prevent or damp the oscillation of the system. When the system attains a pure viscous friction at high speed and viscoelastic friction at low-speed the damping force is called viscous-viscoelastic. Similarly, when the system attains a pure viscoelastic friction at high speed and viscous friction at low-speed the damping force is called viscoelastic- viscous. The damping force is expressed in the form of fractional derivative of position [7, 27-30], with damping constant $c$. Here in this article order of fractional derivative is taken as $q$, which varies continuously.

In the current section, we analyse two suitable cases for linear dampingof fractional continuously variable order mass-spring damper system with single-degree freedom viz.
**Case 1:** Free oscillation with viscoelastic damping
**Case 2**: Forcedoscillation with viscous-viscoelastic or viscoelastic- viscous damping.

## 4.1. Free oscillation with viscoelastic damping

Firstly, we consider the free oscillation of fractional continuously-variable order mass-spring damper system with single-degree freedom. Here the mass " $m$ " displaced from its equilibrium position and then it vibrates freely without any external force $F_0$ .

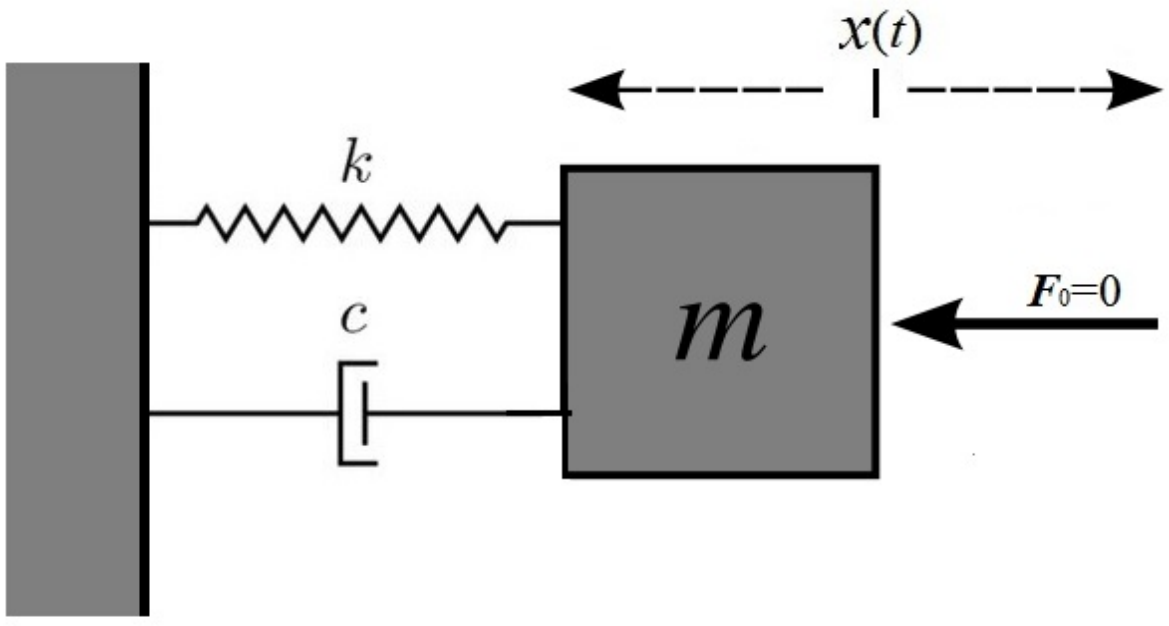

**Fig 1: A mass-spring oscillator under viscoelastic damping when no external force is applied**

For the displaced mass from equilibrium, the system experiences restoring force $F_s$ due to spring constant $k$ , opposing its displacement which is given as:

$$F_s = -kx(t) \qquad (4.1.1)$$

and a damping force $F_d$ ,which is viscoelastic that is described by a fractional continuously-variable order derivative due to viscoelastic of damping coefficient $c$ and given as:

$$F_d = -cD_t^q x(t) \qquad (4.1.2)$$

where $q$ is a continuously variable fractional order viscoelastic oscillator. By the Newton's second law, due to oscillation the free body with mass " $m$ " experiences a total force $F_{Net}$ which is given as

$$F_{Net} = ma \qquad (4.1.3)$$

where " $a$ " denotes the acceleration while the mass oscillates.

$$F_{Net} = mD_t^2 x(t) \qquad (4.1.4)$$

where $a = D_t^2 x(t)$ .

The total force on the body is given as

$$F_{Net} = F_s + F_d + F_0 \qquad (4.1.5)$$

which is equal to

$$mD_t^2 x(t) = -kx(t) - cD_t^q x(t) + F_0 \qquad (4.1.6)$$

We can model the above described problem as a continuous-variable order linear fractional differential equation (FDE) with viscoelastic oscillator, which is described as

$$mD_t^2 x(t) + cD_t^q x(t) + kx(t) = F_0 \qquad (4.1.7)$$

where $q$ is a continuously variable fractional order viscoelastic damper, let us assume that the fractional order $q$ is continuously variable order be defined as

$$q = 0.5 + 0.5\tanh\left(|v|\right) \tag{4.1.8}$$

here $v$ is velocity possessed by the system i.e. $v(t) = \frac{dx(t)}{dt} = x'(t)$ and $0 < v \leq 1$. This continuously variable order is shown in figure-4. In this case, the continuous-variable damper order that is $q$ is maintained to be a function of velocity. For demonstration of a continuously variable fractional order damping we have assumed the function of $q$ varying with velocity in the form (4.1.8). Therefore with this definition of $q$ in equation (4.1.8) at very low speeds the order $q$ tends to 0.5 and the equation of oscillator is with half order damping, that is

$$mD_t^2 x(t) + cD_t^{0.5} x(t) + kx(t) = F_0$$

At high speeds $v = 1$ the fractional order is $q = 0.8807$ and the equation of oscillator is tending towards classical integer order damped oscillator that is described as follows

$$mD_t^2 x(t) + cD_t^{0.8807} + kx(t) = F_0 \qquad mD_t^2 x(t) + cD_t^1 x(t) + kx(t) = F_0$$

Therefore with the oscillation process the fractional order of viscoelastic damping changes continuously with position, time from value half to almost unity, and that also changes the damping order of fractional differential equation (4.1.6). This example we will solve subsequently.

For free oscillation case there is no external force, we take $F_0 = 0$. So the governing eq. (4.1.6) changes to

$$mD_t^2 x(t) + cD_t^q x(t) + kx(t) = 0 \tag{4.1.9}$$

The eq. (4.1.9) can be made continuously-variable order initial value problem by assigning suitable initial conditions. In this case, the continuous-variable-order Initial value problem is well posed for the initial conditions $x(0) = 0$ and $D_t^1\, x(t)\big|_{t=0} \equiv D_t^1 x(0) = x'(0) = 1$.

## 4.2. Forced oscillation with viscous-viscoelastic or viscoelastic- viscous damping.

Consider a fractional continuously-variable order mass-spring damper system with a mass "$m$" oscillating smoothly and repeatedly about its equilibrium position and vibrating freely with external force $F_0$ on a variable viscous- viscoelastic or viscoelastic- viscous path of length $L$. Therefore the generalized damping force that is $F_d = -cD_t^q x(t)$, will be having a continuously variable order $q$ which is depending on position $x(t)$ that is also depending on where at the present instant with-in travel length $L$, the system is positioned. Say we formulate the viscous-viscoelastic damping by $q = \frac{1}{2}\left[1 + \left(\frac{x(t)}{L}\right)^2\right]$, and viscoelastic-viscous damping by $q = \frac{1}{2}\left[1 - \left(\frac{x(t)}{L}\right)^2\right]$, given in (4.2.1). Here $x(t)$ denotes the displacement that is defined for a guide $|x| \leq L$. The guide is the length where the continuous variable damping force exists, and the motion of the mass is with-in the guide's limit. The guide length is thus $2L$, from $x = -L$ to $x = +L$. With $q$ is a viscoelastic-viscous or viscous-viscoelastic type of operator and defined as following

$$q = \begin{cases} \dfrac{1+\left(\frac{x(t)}{L}\right)^2}{2} & \text{if viscous-viscoelastic damping} \\ \dfrac{1-\left(\frac{x(t)}{L}\right)^2}{2} & \text{if viscoelastic-viscous damping} \end{cases} \tag{4.2.1}$$

For the case of viscous-viscoelastic damping, at the beginning position, that is $x(t) = \pm L$ the system starts with a pure viscous friction with order of derivative as $q = 1$, and here $v(\pm L) = x'(\pm L) = 0$ the velocity is low; whereas at high speed that is at $x(t) = 0$ the damping is viscoelastic friction with order of its derivative $q = \frac{1}{2}$. For the case of viscoelastic-viscous damping at very low speeds i.e. at position $x(t) = \pm L$ , the order of derivative is zero, that is there is no friction, whereas at the high speed at $x(t) = 0$, the order of the damping is half. In this article we will solve the case with viscous-viscoelastic damping.

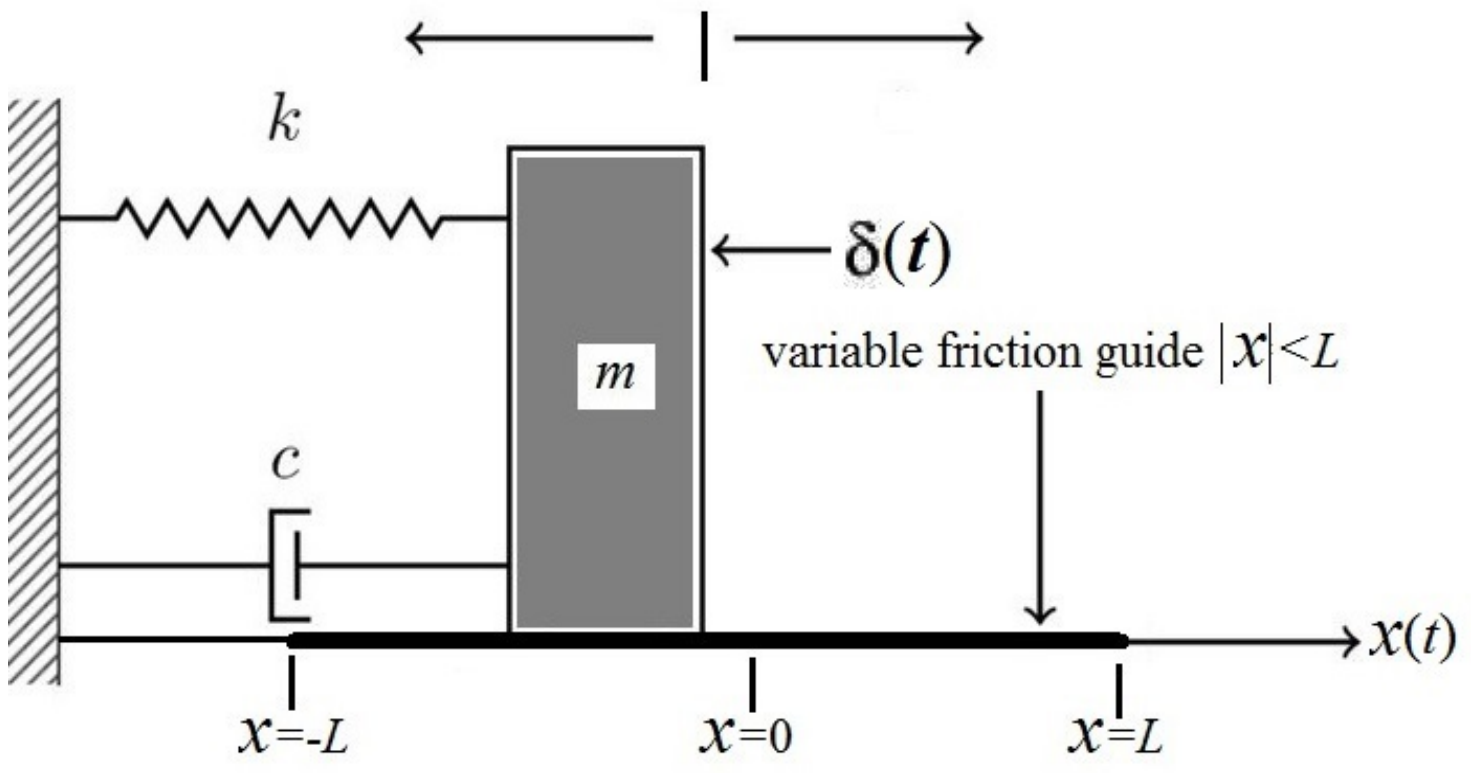

**Fig 2: A mass-spring oscillator sliding on a continuous order guide when external force is applied**

The following figure gives the difference in the two cases

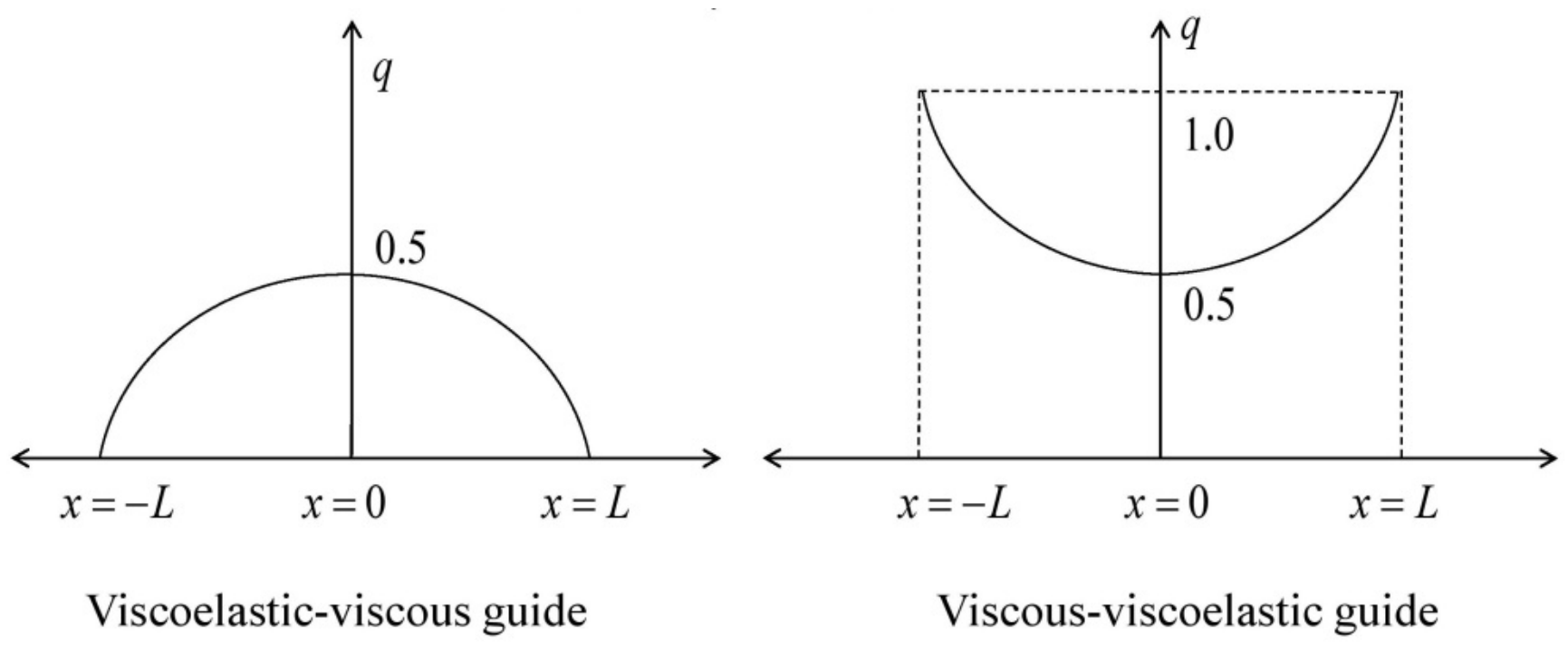

**Fig 3: viscous-viscoelastic and viscoelastic-viscous oscillators**

Suppose body has small impulsive force which is applied externally which is denote as $F_0$ and given as

$$F_0(t) = \delta(t) \tag{4.2.2}$$

where $\delta(t)$ is defined as $\delta(t) = \lim_{\varepsilon \to 0} \frac{1}{2\varepsilon}$ for $-\varepsilon \le t \le \varepsilon$ , else $\delta(t) = 0$ , is Dirac delta function
So the equation of motion is

$$mD_t^2 x(t) + cD_t^q x(t) + kx(t) = \delta(t) \quad (4.2.3)$$

The problem is well defined for $\left|\frac{x(t)}{L}\right| < 1$ and no information about the viscoelastic force is known beyond the limit of the guide. The eq. (4.2.3) can be made continuous-variable order initial value problem by assigning suitable initial conditions. The continuous-order initial value problem is well posed for the initial conditions $x(0) = 0$ and $D_t^1\, x(t)\big|_{t=0} = 0$ . These are rest condition at the start with an impulse force $\delta(t)$ .

# 5. Application of proposed successive iteration method for solution of fractional continuously-variable order mass-spring damper system

The present section includes the analytic solutions (obtained via successive iteration method) for fractional continuously-variable order spring-mass damper systems for free oscillation with viscoelastic damping and force oscillation with viscous-viscoelastic damping. The successive iteration method has been implemented here for finding the analytical solutions for proposed systems.

## 5.1. Implementation of successive iteration method for free oscillation of mass-spring viscoelastic damping system

Consider eq. (4.1.9) with initial conditions, given as

$$x(0) = 0 \text{ and } D_t^1 x(t)\big|_{t=0} = x'(0) = 1 \quad (5.1.1)$$

which tend to the equilibrium states of the proposed dynamic system at the beginning process.
The equation (4.1.9) can be written in the following form

$$D_t^2 x(t) + \frac{c}{m} D_t^q x(t) + \frac{k}{m} x(t) = 0 \quad (5.1.2)$$

here $q$ is the continuously-variable order of a viscous-viscoelastic oscillator, which is defined as in (4.1.8) that we re-write again as following with $t = n\Delta t$ .

$$q = 0.5 + 0.5 \tanh\left(|v|\right)\Big|_{t=n\Delta t} \quad (5.1.3)$$

Here the $q$ changes with the small change with time say $\Delta t$ with velocity and $n \in N$ in (4.1.8), depicted in figure-4. So for each $q$ we have the new solution, which continuously changes throughout the oscillation period. In the successive iteration method that we will use subsequently, we use symbol $D_t x(0)$ , which implies, initial velocity that is $D_t\, x(t)\big|_{t=0} \equiv D_t^1\, x(t)\big|_{t=0} \equiv \frac{dx}{dt}\big|_{t=0} \equiv x'(0)$ .

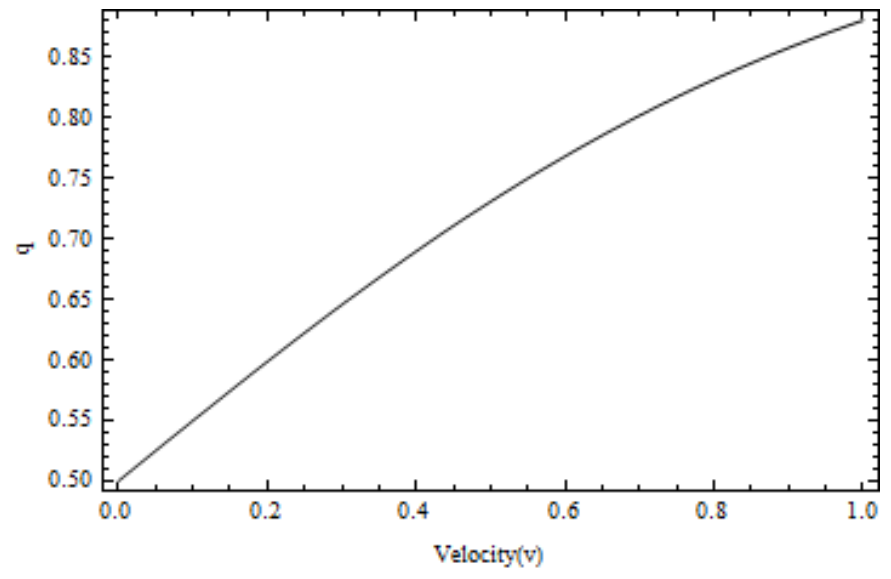

**Fig 4: Figure representing continuously variable viscoelastic oscillator**

By successive iterative method eq. (5.1.2) can be written as

$$x(t) = x(0) + tD_t x(0) - \frac{k}{m} L^{-1} x(t) - \frac{c}{m} L^{-1} D_t^q x(t) \tag{5.1.4}$$

Here the inverse linear operator is taken as $L^{-1} = D_t^{-2}$, that is 2-fold definite integration operation from $0$ to $t$ .The eq. (5.1.4) can be rewritten as

$$x(t) = x(0) + tD_t x(0) - \frac{k}{m} D_t^{-2} x(t) - \frac{c}{m} D_t^{-2}\left(D_t^q x(t)\right)$$

$$= x(0) + tD_t x(0) - \frac{k}{m} D_t^{-2} x(t) - \frac{c}{m} D_t^{-2+q} x(t) \tag{5.1.5}$$

By using initial conditions (5.1.1), we can calculate initially for first iteration $q$ as

$$q = 0.5 + 0.5 \tanh\left(|v|\right)\Big|_{t=0}$$

$$= 0.5 + 0.5 \tanh(1) = 0.8807 \tag{5.1.6}$$

**First iteration:**

Therefore equation (5.1.5) can be written as

$$x(t) = x(0) + tD_t x(0) - \frac{k}{m} D_t^{-2} x(t) - \frac{c}{m} D_t^{-1.1192} x(t) \tag{5.1.7}$$

We use the Euler's formula for fractional integration of a power-function that is $D_t^{-\alpha}\left[t^m\right] = \frac{\Gamma(m+1)}{\Gamma(m+\alpha+1)} t^{m+\alpha}$ .So by successive iteration method, we have the following iterations

$$x_0(t) = x(0) + tD_t x(0) = t$$

$$x_1(t) = -\frac{k}{m} D_t^{-2} x_0(t) - \frac{c}{m} D_t^{-1.1192} x_0(t)$$

$$= -\frac{k}{m}\frac{t^3}{\Gamma(4)} - \frac{c}{m}\frac{t^{2.1192}}{\Gamma(3.1192)};$$

$$x_2(t) = -\frac{k}{m} D_t^{-2} x_1(t) - \frac{c}{m} D_t^{-1.1192} x_1(t)$$

$$= \frac{k^2}{m^2}\frac{t^5}{\Gamma(6)} + \frac{2kc}{m^2}\frac{t^{4.1192}}{\Gamma(5.1192)} + \frac{c^2}{m^2}\frac{t^{3.2384}}{\Gamma(4.2384)};$$

$$x_3(t) = -\frac{k}{m} D_t^{-2} x_2(t) - \frac{c}{m} D_t^{-1.1192} x_2(t)$$

$$= -\frac{k^3}{m^3}\frac{t^7}{\Gamma(8)} - \frac{3k^2c}{m^3}\frac{t^{6.1192}}{\Gamma(7.1192)} - \frac{3kc^2}{m^3}\frac{t^{5.3284}}{\Gamma(6.2384)} - \frac{c^3}{m^3}\frac{t^{4.3576}}{\Gamma(5.3576)};$$

and so on. So the solution of eq. (4.1.9) for $q = 0.880797$ is given as

$$x(t) = x_0(t) + x_1(t) + x_2(t) + x_3(t)$$

$$= t - \frac{k}{m}\frac{t^3}{\Gamma(4)} - \frac{c}{m}\frac{t^{2.1192}}{\Gamma(3.1192)} + \frac{k^2}{m^2}\frac{t^5}{\Gamma(6)} + \frac{2kc}{m^2}\frac{t^{4.1192}}{\Gamma(5.1192)} + \frac{c^2}{m^2}\frac{t^{3.2384}}{\Gamma(4.2384)}$$

$$-\frac{k^3}{m^3}\frac{t^7}{\Gamma(8)} - \frac{3k^2c}{m^3}\frac{t^{6.1192}}{\Gamma(7.1192)} - \frac{3kc^2}{m^3}\frac{t^{5.2384}}{\Gamma(6.2384)} - \frac{c^3}{m^3}\frac{t^{4.3576}}{\Gamma(5.3576)} \tag{5.1.8}$$

**Second iteration:**

By differentiation we obtain from (5.1.8) $v(t) = x'(t)$ that is velocity. Take the values $\frac{k}{m} = \omega_n^2$, $\frac{c}{m} = 2\eta\omega_n^{3/2}$, $\omega_n = 2$, $\eta = 0.5$ in eq. (5.1.8); this substitution is done to make the equation of motion in similar lines that of a classical damped oscillator equation, Elshehaway et al [10], Gomez et al [14], Bagley et al [20], Saha Ray et al [23], Torvic et al [30].We have also solved for other values of $\eta$ as we report in graphical plots in subsequent section.With this we have by putting $\Delta t = 0.01$, that is for the next time-step $n = 1$, $t = 0.01$ we obtain $x(0.01) = 0.0099267$ and $x'(0.01) = v(0.01) = 0.984444$

From the expression (4.1.8) of variable order of viscoelastic element, we obtain the next value of the variable order that is

$$q = 0.5 + 0.5\tanh\left(\left|v\right|\right)\Big|_{t=0.01}$$
$$= 0.5 + 0.5\tanh(0.984444) = 0.877492$$

Here we note that initially at $t = 0$ we started with order of visco-elastic element with fractional order $q = 0.8807$ and in next iteration, for $t = 0.01$ with change in time, we have the new value of fractional order of viscoelastic damper that is $q = 0.877492$ .So eq. (5.1.5) becomes with changed fractional order of damping as described below

$$x(t) = x(0) + tD_t x(0) - \frac{k}{m} D_t^{-2} x(t) - \frac{c}{m} D_t^{-1.122508} x(t) \quad (5.1.9)$$

So by successive iteration method, we have

$$x_0(t) = x(0) + tD_t x(0) = t$$

$$x_1(t) = -\frac{k}{m} D_t^{-2} x_0(t) - \frac{c}{m} D_t^{-1.122508} x_0(t)$$
$$= -\frac{k}{m}\frac{t^3}{\Gamma(4)} - \frac{c}{m}\frac{t^{2.122508}}{\Gamma(3.122508)};$$

$$x_2(t) = -\frac{k}{m} D_t^{-2} x_1(t) - \frac{c}{m} D_t^{-1.122508} x_1(t)$$
$$= \frac{k^2}{m^2}\frac{t^5}{\Gamma(6)} + \frac{2kc}{m^2}\frac{t^{4.122508}}{\Gamma(5.122508)} + \frac{c^2}{m^2}\frac{t^{3.45016}}{\Gamma(4.45016)};$$

$$x_3(t) = -\frac{k}{m} D_t^{-2} x_2(t) - \frac{c}{m} D_t^{-1.122508} x_2(t)$$
$$= -\frac{k^3}{m^3}\frac{t^7}{\Gamma(8)} - \frac{3k^2c}{m^3}\frac{t^{6.122508}}{\Gamma(7.122508)} - \frac{3kc^2}{m^3}\frac{t^{5.45016}}{\Gamma(6.45016)} - \frac{c^3}{m^3}\frac{t^{4.67524}}{\Gamma(5.67524)}$$

and so on. So the solution of eq. (4.1.9) for $q = 0.877492$ is given as

$$x(t) = x_0(t) + x_1(t) + x_2(t) + x_3(t)$$
$$= t - \frac{k}{m}\frac{t^3}{\Gamma(4)} - \frac{c}{m}\frac{t^{2.122508}}{\Gamma(3.122508)} + \frac{k^2}{m^2}\frac{t^5}{\Gamma(6)} + \frac{2kc}{m^2}\frac{t^{4.122508}}{\Gamma(5.122508)} + \frac{c^2}{m^2}\frac{t^{3.45016}}{\Gamma(4.45016)}$$
$$- \frac{k^3}{m^3}\frac{t^7}{\Gamma(8)} - \frac{3k^2c}{m^3}\frac{t^{6.122508}}{\Gamma(7.122508)} - \frac{3kc^2}{m^3}\frac{t^{5.45016}}{\Gamma(6.45016)} - \frac{c^3}{m^3}\frac{t^{4.67524}}{\Gamma(5.67524)} \quad (5.1.10)$$

Similarly by taking $t = n\Delta t$ and using eq. (5.1.3), the fractional order $q$ for next all iteration can be calculated. Here $\Delta t = 0.01$ i.e. the time step is 0.01 seconds and $n = 2,3,4,\ldots$.

By generalizing the solution by successive iteration method, we have the following, where $q$ is time dependent too, call that $q_{n\Delta t}$

$$x_0(t) = x(0) + tD_t x(0) = t;$$

$$x_1(t) = -\frac{k}{m} D_t^{-2} x_0(t) - \frac{c}{m} D_t^{-2+q_{n\Delta t}} x_0(t)$$
$$= -\frac{k}{m}\frac{t^3}{\Gamma(4)} - \frac{c}{m}\frac{t^{3-q_{n\Delta t}}}{\Gamma(4-q_{n\Delta t})};$$

$$x_2(t) = -\frac{k}{m} D_t^{-2} x_1(t) - \frac{c}{m} D_t^{-2+q_{n\Delta t}} x_1(t)$$
$$= \frac{k^2}{m^2}\frac{t^5}{\Gamma(6)} + \frac{2kc}{m^2}\frac{t^{5-q_{n\Delta t}}}{\Gamma(6-q_{n\Delta t})} + \frac{c^2}{m^2}\frac{t^{5-2q_{n\Delta t}}}{\Gamma(6-2q_{n\Delta t})};$$

$$x_3(t) = -\frac{k}{m} D_t^{-2} x_2(t) - \frac{c}{m} D_t^{-2+q_{n\Delta t}} x_2(t)$$

$$= -\frac{k^3}{m^3}\frac{t^7}{\Gamma(8)} - \frac{3k^2c}{m^3}\frac{t^{7-q_{n\Delta t}}}{\Gamma(8-q_{n\Delta t})} - \frac{3kc^2}{m^3}\frac{t^{7-2q_{n\Delta t}}}{\Gamma(8-2q_{n\Delta t})} - \frac{c^3}{m^3}\frac{t^{7-3q_{n\Delta t}}}{\Gamma(8-3q_{n\Delta t})}$$

And so on.

The total solution is therefore

$$x(t) = x_0(t) + x_1(t) + x_2(t) + x_3(t).....$$

$$= t - \frac{k}{m}\frac{t^3}{\Gamma(4)} - \frac{c}{m}\frac{t^{3-q_{n\Delta t}}}{\Gamma(4-q_{n\Delta t})} + \frac{k^2}{m^2}\frac{t^5}{\Gamma(6)} + \frac{2kc}{m^2}\frac{t^{5-q_{n\Delta t}}}{\Gamma(6-q_{n\Delta t})} + \frac{c^2}{m^2}\frac{t^{5-2q_{n\Delta t}}}{\Gamma(6-2q_{n\Delta t})}$$

$$- \frac{k^3}{m^3}\frac{t^7}{\Gamma(8)} - \frac{3k^2c}{m^3}\frac{t^{7-q_{n\Delta t}}}{\Gamma(8-q_{n\Delta t})} - \frac{3kc^2}{m^3}\frac{t^{7-2q_{n\Delta t}}}{\Gamma(8-2q_{n\Delta t})} - \frac{c^3}{m^3}\frac{t^{7-3q_{n\Delta t}}}{\Gamma(8-3q_{n\Delta t})} + ........$$

The value of $q$ is variable and that in this case depends on velocity or $x'(t)$, and also with $t$ given by expression (4.1.8). With rearrangement in above series solution we can write the above series as following compact expression as following

$$x(t) = \sum_{r=0}^{\infty} \frac{\left(-\frac{k}{m}\right)^r}{r!} t^{2r+1} \sum_{j=0}^{\infty} \frac{\left(-\frac{c}{m}\right)^j (j+r)!\, t^{(2-q_{n\Delta t})j}}{j!\,\Gamma((2-q_{n\Delta t})j + 2r + 2)}$$

$$= \sum_{r=0}^{\infty} \frac{\left(-\frac{k}{m}\right)^r}{r!} t^{2r+1} E^{(r)}_{2-q_{n\Delta t},\frac{r}{2}+2}\left(\left(-\frac{c}{m}\right) t^{(2-q_{n\Delta t})}\right)$$

$$= \sum_{r=0}^{\infty} \frac{\left(-\omega_n^2\right)^r}{r!} t^{2r+1} E^{(r)}_{2-q_{n\Delta t},\frac{r}{2}+2}\left(\left(-2\eta\omega_n^{3/2}\right) t^{(2-q_{n\Delta t})}\right) \qquad (5.1.11)$$

where $E_{\lambda,\mu}$ defines the Mittag-Leffler function in two parameter is

$$E_{\lambda,\mu}(y) = \sum_{j=0}^{\infty} \frac{y^j}{\Gamma(\lambda j + \mu)} \text{ for } \lambda > 0 \text{ and } \mu > 0$$

and the $r-$th derivative of the two parameter Mittag-Leffler function is defined as

$$E^{(r)}_{\lambda,\mu}(y) = \frac{d^r}{dy^r} E_{\lambda,\mu} = \sum_{j=0}^{\infty} \frac{(j+r)!\, y^j}{j!\,\Gamma(\lambda j + \lambda r + \mu)}, \ (r = 0,1,2,...)$$

The choice of time step of 0.01 is for convenience. Ideally it should be as small as possible. A smaller value of time step that 0.01 gives a very large time to obtain the the solution in computer; and a larger value of time step gives inaccurate results. The idea is to simulate the results for a continuously variable order; and we found the 0.01 time-step to be convenient for our 600 steps iteration, which are plotted in the graphs.

## 5.2. Application of successive iteration method for forced oscillation of spring-mass viscous-viscoelastic damping system

Let us consider eq. (4.2.3) with the homogenous initial conditions

$$x(0) = 0 \text{ and } D_t^1 x(t)\Big|_{t=0} = 0 \qquad (5.2.1)$$

which is at the equilibrium states of the dynamic system at the beginning process. The equation (4.2.3) can be written in the following form

$$D_t^2 x(t) + \frac{c}{m} D_t^q x(t) + \frac{k}{m} x(t) = \frac{\delta(t)}{m} \qquad (5.2.2)$$

here $q$ is the fractional continuously-variable order viscous-viscoelastic oscillator, which is defined as

$$q=\frac{1+\left(\frac{x(t)}{L}\right)^2\Big|_{t=n\Delta t}}{2} \tag{5.2.3}$$

Here the $q$ changes with the small change in time say $\Delta t$ and $n\in N$ in (5.2.3). So for each $t=n\Delta t$ we have the new fractional differential equation and its solution, which continuously changes throughout the period of oscillation. By successive iterative method equation (5.2.2) can be written as

$$x(t)=x(0)+tD_t x(0)+L^{-1}\frac{\delta(t)}{m}-\frac{k}{m}L^{-1}x(t)-\frac{c}{m}L^{-1}D_t^q x(t) \tag{5.2.4}$$

Here the inverse linear operator is taken as $L^{-1}=D_t^{-2}$ .The eq. (5.2.4) can be rewritten as

$$x(t)=x(0)+tD_t x(0)+D_t^{-2}\frac{\delta(t)}{m}-\frac{k}{m}D_t^{-2}x(t)-\frac{c}{m}D_t^{-2}\left(D_t^q x(t)\right)$$

$$=x(0)+tD_t x(0)+D_t^{-2}\frac{\delta(t)}{m}-\frac{k}{m}D_t^{-2}x(t)-\frac{c}{m}D_t^{-2+q}x(t) \tag{5.2.5}$$

By using initial conditions (5.2.1) and by taking unit length of regime i.e. $L=1$, we can calculate initially for first iteration $q$ as

$$q=\frac{1+\left(\frac{x(0)}{L}\right)^2}{2}=\frac{1}{2} \tag{5.2.6}$$

**First iteration:**

Therefore equation (5.2.5) can be written as

$$x(t)=x(0)+tD_t x(0)+D_t^{-2}\frac{\delta(t)}{m}-\frac{k}{m}D_t^{-2}x(t)-\frac{c}{m}D_t^{-1.5}x(t) \tag{5.2.7}$$

So by successive iteration method, we have

$$x_0(t)=x(0)+tD_t x(0)+D_t^{-2}\frac{\delta(t)}{m}=\frac{t}{m};$$

$$x_1(t)=-\frac{k}{m}D_t^{-2}x_0(t)-\frac{c}{m}D_t^{-1.5}x_0(t)$$

$$=-\frac{k}{m^2}\frac{t^3}{\Gamma(4)}-\frac{c}{m^2}\frac{t^{2.5}}{\Gamma(3.5)};$$

$$x_2(t)=-\frac{k}{m}D_t^{-2}x_1(t)-\frac{c}{m}D_t^{-1.5}x_1(t)$$

$$=\frac{k^2}{m^3}\frac{t^5}{\Gamma(6)}+\frac{2kc}{m^3}\frac{t^{4.5}}{\Gamma(5.5)}+\frac{c^2}{m^3}\frac{t^4}{\Gamma(5)};$$

$$x_3(t)=-\frac{k}{m}D_t^{-2}x_2(t)-\frac{c}{m}D_t^{-1.5}x_2(t)$$

$$=-\frac{k^3}{m^4}\frac{t^7}{\Gamma(8)}-\frac{3k^2c}{m^4}\frac{t^{6.5}}{\Gamma(7.5)}-\frac{3kc^2}{m^4}\frac{t^6}{\Gamma(7)}-\frac{c^3}{m^4}\frac{t^{5.5}}{\Gamma(6.5)}$$

and so on. So the solution of eq. (4.2.3) for $q(x(t))\big|_{t=0}=\frac{1}{2}$ is given as

$$x(t)=x_0(t)+x_1(t)+x_2(t)+x_3(t)$$

$$=\frac{t}{m}-\frac{k}{m^2}\frac{t^3}{\Gamma(4)}-\frac{c}{m^2}\frac{t^{2.5}}{\Gamma(3.5)}+\frac{k^2}{m^3}\frac{t^5}{\Gamma(6)}+\frac{2kc}{m^3}\frac{t^{4.5}}{\Gamma(5.5)}+\frac{c^2}{m^3}\frac{t^4}{\Gamma(5)}$$

$$-\frac{k^3}{m^4}\frac{t^7}{\Gamma(8)}-\frac{3k^2c}{m^4}\frac{t^{6.5}}{\Gamma(7.5)}-\frac{3kc^2}{m^4}\frac{t^6}{\Gamma(7)}-\frac{c^3}{m^4}\frac{t^{5.5}}{\Gamma(6.5)} \tag{5.2.8}$$

**Second iteration:**

By putting $\Delta t=0.01$, that is for $n=1$, and taking $m=1$, $\frac{k}{m}=\omega_n^2$, $\frac{c}{m}=2\eta\omega_n^{3/2}$, $\omega_n=2$ and $\eta=0.5$ in equation (5.2.8), we have $x(0.01)=0.0100078$. We have also solved for other values of $\eta$ as we report in graphical plots in subsequent section. By taking unit length i.e. $L=1$ and $x(0.01)=0.0100078$, we find $q$ for second iteration and calculated as follows

$$q=\frac{1+\left(\frac{x(0.01)}{L}\right)^2}{2}=0.50005$$

The equation (5.2.5) can be written as

$$x(t)=x(0)+tD_t x(0)+D_t^{-2}\frac{\delta(t)}{m}-\frac{k}{m}D_t^{-2}x(t)-\frac{c}{m}D_t^{-1.49995}x(t) \qquad (5.2.9)$$

So by successive iteration method, we have

$$x_0(t)=x(0)+tD_t x(0)+D_t^{-2}\frac{\delta(t)}{m}=\frac{t}{m};$$

$$x_1(t)=-\frac{k}{m}D_t^{-2}x_0(t)-\frac{c}{m}D_t^{-1.49995}x_0(t)$$

$$=-\frac{k}{m^2}\frac{t^3}{\Gamma(4)}-\frac{c}{m^2}\frac{t^{2.49995}}{\Gamma(3.49995)};$$

$$x_2(t)=-\frac{k}{m}D_t^{-2}x_1(t)-\frac{c}{m}D_t^{-1.49995}x_1(t)$$

$$=\frac{k^2}{m^3}\frac{t^5}{\Gamma(6)}+\frac{2kc}{m^3}\frac{t^{4.49995}}{\Gamma(5.49995)}+\frac{c^2}{m^3}\frac{t^{3.9999}}{\Gamma(4.9999)};$$

$$x_3(t)=-\frac{k}{m}D_t^{-2}x_2(t)-\frac{c}{m}D_t^{-1.49995}x_2(t)$$

$$=-\frac{k^3}{m^4}\frac{t^7}{\Gamma(8)}-\frac{3k^2c}{m^4}\frac{t^{6.49995}}{\Gamma(7.49995)}-\frac{3kc^2}{m^4}\frac{t^{5.9999}}{\Gamma(6.9999)}-\frac{c^3}{m^4}\frac{t^{5.49985}}{\Gamma(6.49985)}$$

and so on. So the solution of eq. (4.2.3) for $q=0.50005$ is given as

$$x(t)=x_0(t)+x_1(t)+x_2(t)+x_3(t)$$

$$=\frac{t}{m}-\frac{k}{m^2}\frac{t^3}{\Gamma(4)}-\frac{c}{m^2}\frac{t^{2.49995}}{\Gamma(3.49995)}+\frac{k^2}{m^3}\frac{t^5}{\Gamma(6)}+\frac{2kc}{m^3}\frac{t^{4.49995}}{\Gamma(5.49995)}+\frac{c^2}{m^3}\frac{t^{3.9999}}{\Gamma(4.9999)}$$

$$-\frac{k^3}{m^4}\frac{t^7}{\Gamma(8)}-\frac{3k^2c}{m^4}\frac{t^{6.49995}}{\Gamma(7.49995)}-\frac{3kc^2}{m^4}\frac{t^{5.9999}}{\Gamma(6.9999)}-\frac{c^3}{m^4}\frac{t^{5.49985}}{\Gamma(6.49985)} \qquad (5.2.10)$$

Similarly by taking $t=n\Delta t$, $t=n\Delta t$ and using eq. (5.2.3), the fractional order $q$ for next all iteration can be calculated. Here $\Delta t=0.01$ i.e. the time step is 0.01 seconds and $n=2,3,4,\dots$.

By generalizing the solution by successive iteration method (as carried out in the previous section), we have the following

$$x_0(t)=x(0)+tD_t x(0)+D_t^{-2}\frac{\delta(t)}{m}=\frac{t}{m};$$

$$x_1(t)=-\frac{k}{m}D_t^{-2}x_0(t)-\frac{c}{m}D_t^{-2+q_{n\Delta t}}x_0(t)$$

$$=-\frac{k}{m^2}\frac{t^3}{\Gamma(4)}-\frac{c}{m^2}\frac{t^{3-q_{n\Delta t}}}{\Gamma(4-q_{n\Delta t})};$$

$$x_2(t)=-\frac{k}{m}D_t^{-2}x_1(t)-\frac{c}{m}D_t^{-2+q_{n\Delta t}}x_1(t)$$

$$=\frac{k^2}{m^3}\frac{t^5}{\Gamma(6)}+\frac{2kc}{m^3}\frac{t^{5-q_{n\Delta t}}}{\Gamma(6-q_{n\Delta t})}+\frac{c^2}{m^3}\frac{t^{5-2q_{n\Delta t}}}{\Gamma(6-2q_{n\Delta t})};$$

$$x_3(t)=-\frac{k}{m}D_t^{-2}x_2(t)-\frac{c}{m}D_t^{-2+q_{n\Delta t}}x_2(t)$$

$$= -\frac{k^3}{m^4}\frac{t^7}{\Gamma(8)} - \frac{3k^2c}{m^4}\frac{t^{7-q_{n\Delta t}}}{\Gamma(8-q_{n\Delta t})} - \frac{3kc^2}{m^4}\frac{t^{7-2q_{n\Delta t}}}{\Gamma(8-2q_{n\Delta t})} - \frac{c^3}{m^4}\frac{t^{7-3q_{n\Delta t}}}{\Gamma(8-3q_{n\Delta t})}$$

and so on.
The total solution is

$$x(t) = x_0(t) + x_1(t) + x_2(t) + x_3(t) + ...$$

$$= \frac{t}{m} - \frac{k}{m^2}\frac{t^3}{\Gamma(4)} - \frac{c}{m^2}\frac{t^{3-q_{n\Delta t}}}{\Gamma(4-q_{n\Delta t})} + \frac{k^2}{m^3}\frac{t^5}{\Gamma(6)} + \frac{2kc}{m^3}\frac{t^{5-q_{n\Delta t}}}{\Gamma(6-q_{n\Delta t})} + \frac{c^2}{m^3}\frac{t^{5-2q_{n\Delta t}}}{\Gamma(6-2q_{n\Delta t})}$$

$$-\frac{k^3}{m^4}\frac{t^7}{\Gamma(8)} - \frac{3k^2c}{m^4}\frac{t^{7-q_{n\Delta t}}}{\Gamma(8-q_{n\Delta t})} - \frac{3kc^2}{m^4}\frac{t^{7-2q_{n\Delta t}}}{\Gamma(8-2q_{n\Delta t})} - \frac{c^3}{m^4}\frac{t^{7-3q_{n\Delta t}}}{\Gamma(8-3q_{n\Delta t})} + .......$$

The value of $q$ is variable and that in this case depends on position $x(t)$ given by expression (5.2.3). With rearrangement in above series solution we can write the above series as following compact expression as following

$$x(t) = \frac{1}{m}\sum_{r=0}^{\infty}\frac{\left(-\frac{k}{m}\right)^r}{r!}t^{2r+1}\sum_{j=0}^{\infty}\frac{\left(-\frac{c}{m}\right)^j (j+r)!t^{(2-q_{n\Delta t})j}}{j!\Gamma((2-q_{n\Delta t})j+2r+2)}$$

$$= \frac{1}{m}\sum_{r=0}^{\infty}\frac{\left(-\frac{k}{m}\right)^r}{r!}t^{2r+1}E^{(r)}_{2-q_{n\Delta t},\frac{r}{2}+2}\left(\left(-\frac{c}{m}\right)t^{(2-q_{n\Delta t})}\right)$$

$$= \frac{1}{m}\sum_{r=0}^{\infty}\frac{\left(-\omega_n^2\right)^r}{r!}t^{2r+1}E^{(r)}_{2-q_{n\Delta t},\frac{r}{2}+2}\left(\left(-2\eta\omega_n^{3/2}\right)t^{(2-q_{n\Delta t})}\right) \quad (5.2.11)$$

where $E_{\lambda,\mu}$ defines the Mittag-Leffler function in two parameter, and the $r-$derivative of two-parameter Mittag-Leffler function isdefined as

$$E^{(r)}_{\lambda,\mu}(y) = \frac{d^r}{dy^r}E_{\lambda,\mu} = \sum_{j=0}^{\infty}\frac{(j+r)!y^j}{j!\Gamma(\lambda j+\lambda r+\mu)}, \quad (r=0,1,2,...)$$

## 6. Numerical simulations and discussion

The solution to the oscillator problem with continuously variable damping order $q$, defined in first case as viscoelastic damping via expression (5.1.3) and in second case as viscous-viscoelastic damping via expression (5.2.3) governed by fractional differential equations $D_t^2x(t) + \frac{c}{m}D_t^q x(t) + \frac{k}{m}x(t) = 0$ and $D_t^2x(t) + \frac{c}{m}D_t^q x(t) + \frac{k}{m}x(t) = \delta(t)$ respectively, gives same type of solution in compact form expressed in equations (5.1.11) and (5.2.11) respectively. In the first case though we are not having external forcing function, but the initial condition at $t=0$ is $x(0)=0$ and $D_t^1x(t)\big|_{t=0} = x'(0) = 1$, that is the system has initial velocity of unity magnitude. In the second case we have Dirac-delta function as forcing function that is applied at $t=0$ with system initially at rest i.e. at $t=0$ we have $x(0)=0$ and $D_t x(t)\big|_{t=0} = x'(0) = 0$. In the second case the forcing function starts the oscillations. Both are cases of free running damped oscillator, with continuously varying order of damping, that is $q$ throughout the oscillation period.

In the present section, the displacement-time graphs have been presented for fractional continuously variable order mass-spring damper systems for free oscillation with viscoelastic damping and forced oscillation with viscous-viscoelastic damping. Thus the two fractional differential equations with continuous variable order $q$ are following; with the initial conditions stated as above sections.

$$D_t^2x(t) + 2\eta\omega_n^{3/2}D_t^q x(t) + \omega_n^2 x(t) = 0 \qquad D_t^2x(t) + 2\eta\omega_n^{3/2}D_t^q x(t) + \omega_n^2 x(t) = \delta(t)$$

Here $\omega_n$ is called the natural or angular frequency and $\eta$ is called the damping ratio of the system.

## 6.1. Numerical simulation of fractional continuously-variable order mass-spring damper system for free oscillation with viscoelastic damping

For case-one, the body oscillates without implementation of any external forces with the continuous change of the fractional continuously variable order viscoelastic damper $q$, with $q(x'(t)) = 0.5 + 0.5\tanh(|x'(t)|)$. Here we have taken $0 \le t \le 6$ and 600 iterations for plotting the time response of the fractional continuous order damping system with displacement $x(t)$ and velocity $v(t)$. Also in the solution (5.1.11), we have taken mass $m = 1$ unit, natural or angular frequency $\omega_n = 2$ rad/sec and the damping ratio $\eta = 0.05, 0.5, 1, \sqrt{\pi}$ .

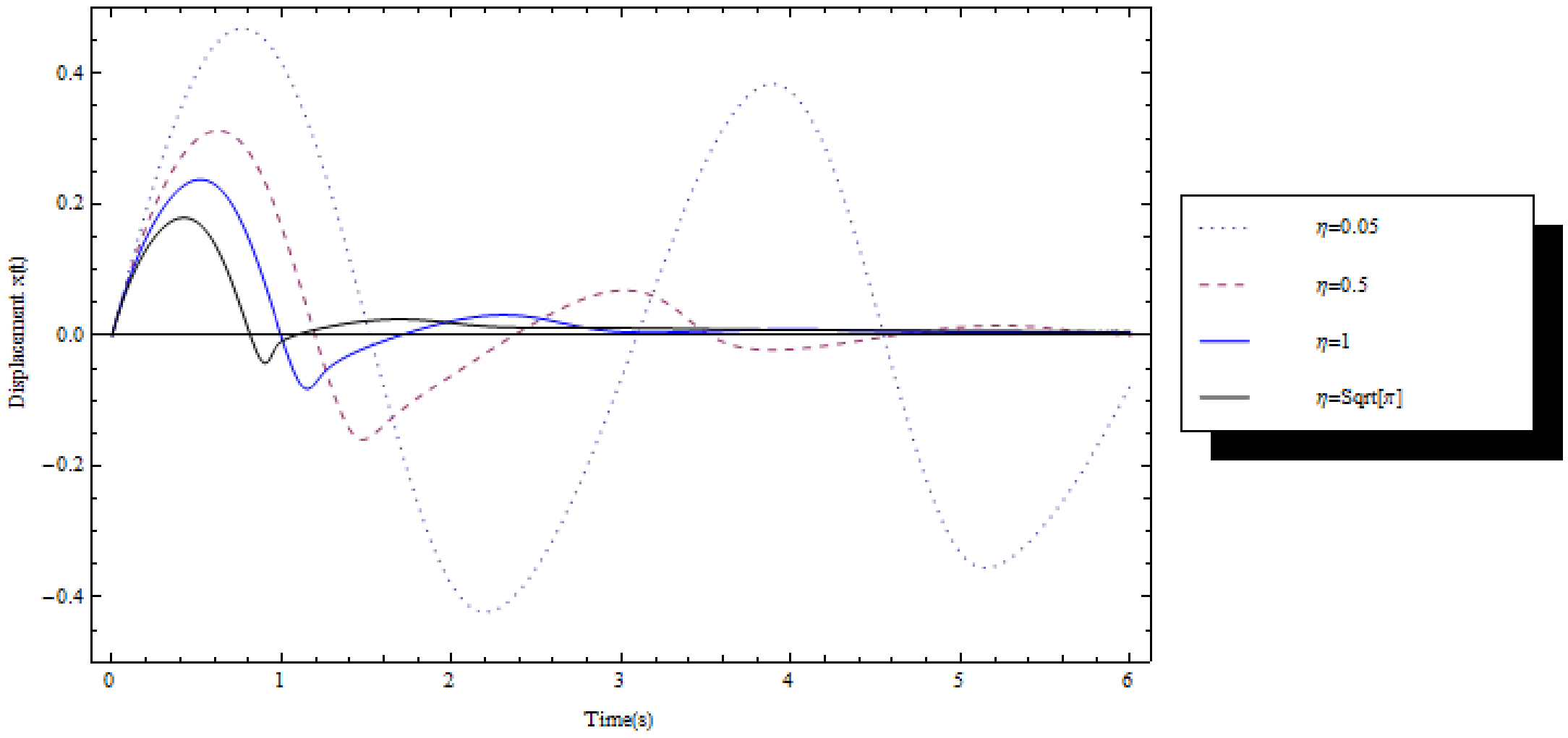

**Fig 5: The displacement-time graph for fractional continuous order spring-mass damper model for free oscillation with viscoelastic damping plots**

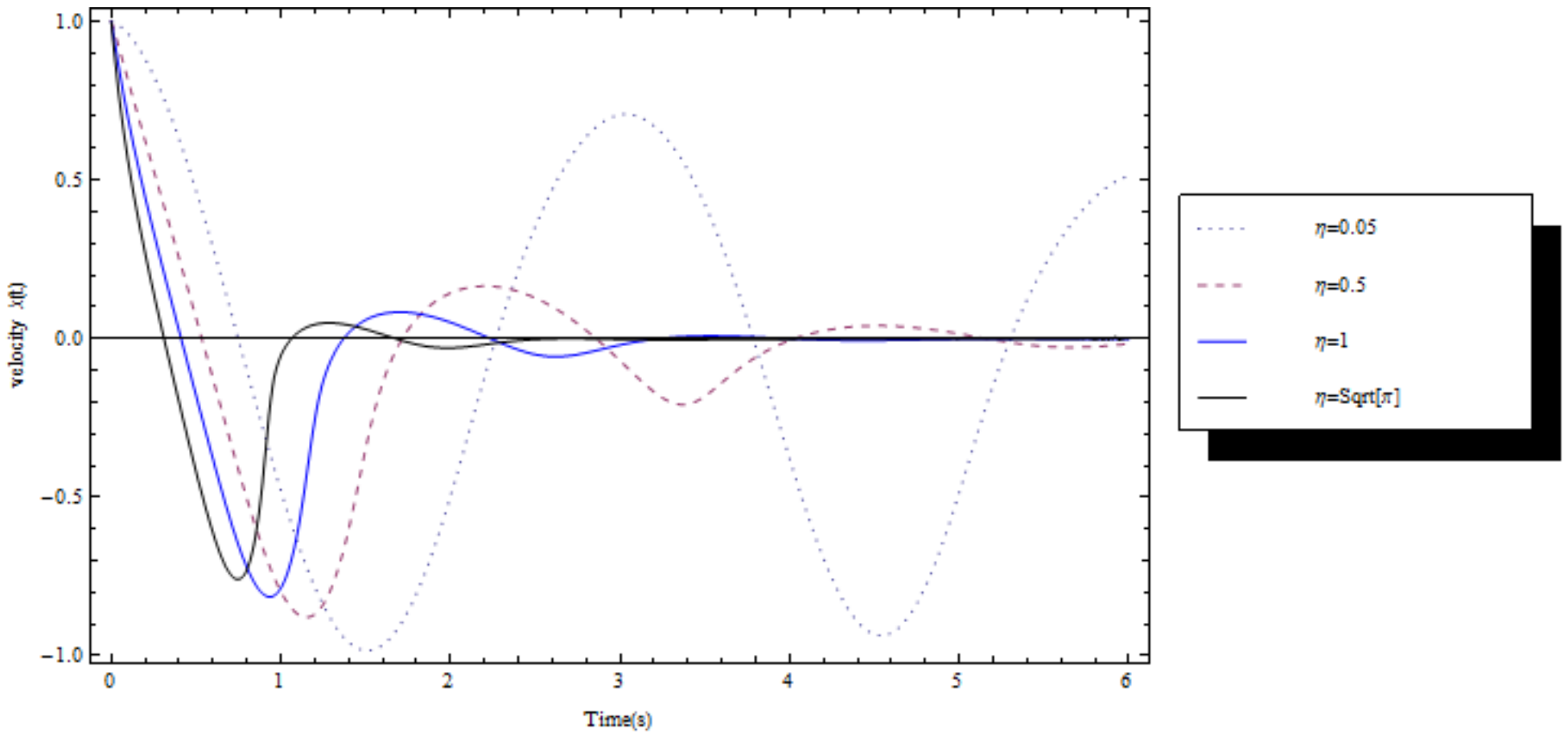

**Fig 6: The velocity-time graph for fractional continuous order spring-mass damper model for free oscillation with viscoelastic damping plots**

## 6.2. Numerical simulation of fractional continuously-variable order mass-spring damper system for forced oscillation with viscous-viscoelastic damping

It is important to mention here that, in system in case-two, oscillates with the small external impulse force $\delta(t)$ with the continuous change of the fractional order viscous-viscoelastic oscillator $q$, with $q\left(x(t)\right)=\frac{1}{2}\left[1+\left(\frac{x(t)}{L}\right)^2\right]$ which has range between 0.5 to 1 i.e. $0.5 \le q \le 1$. Here we have taken $0 \le t \le 6$ and 600 iterations for plotting the time response of the fractional continuous order damping system with displacement $x(t)$. Also in the solution (5.2.11), we have taken mass $m=1$ unit, natural or angular frequency $\omega_n = 2$ rad/sec and the damping ratio $\eta = 0.05, 0.5, 1, \sqrt{\pi}$ .

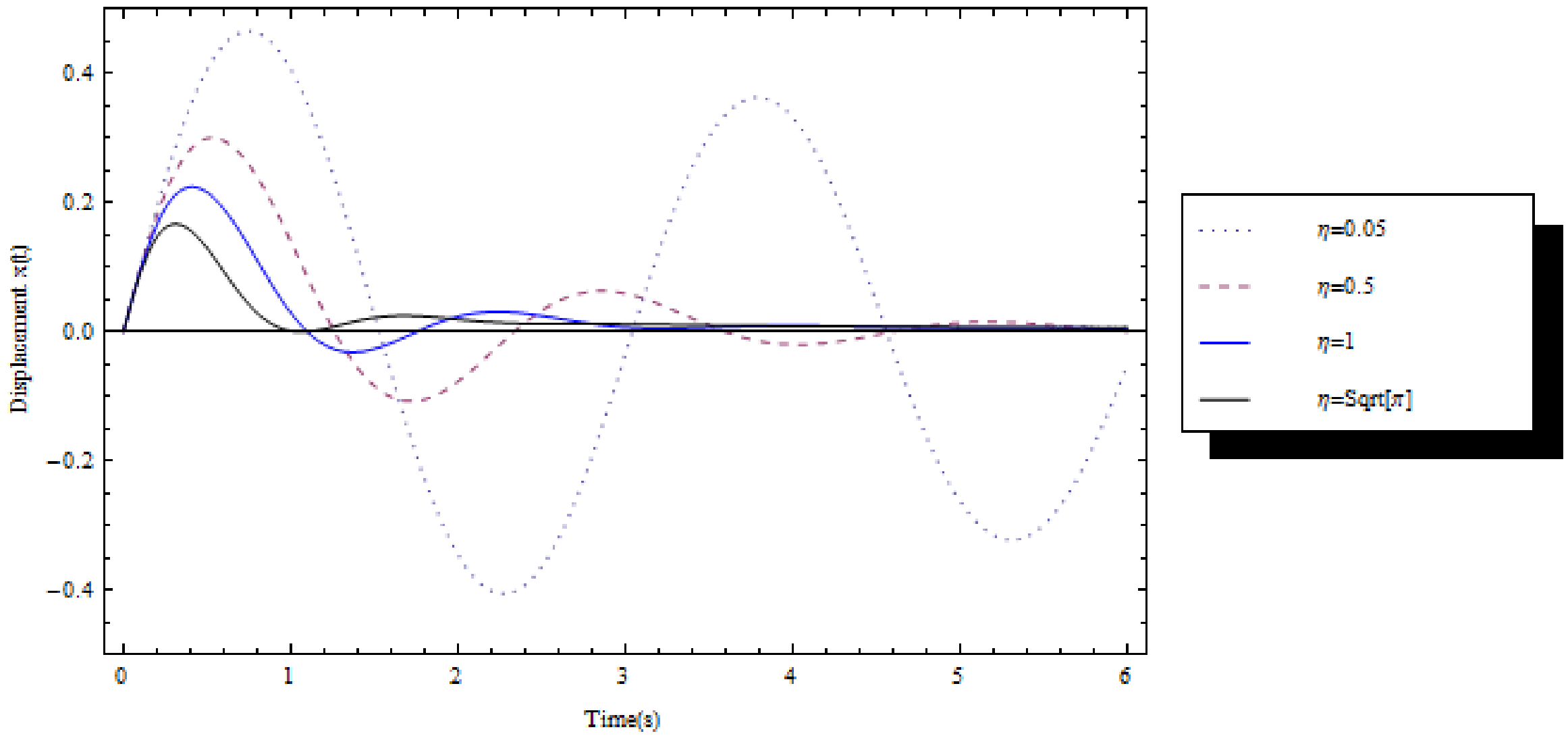

**Fig 7: The displacement-time graph for fractional continuous order mass-spring damper model for forced oscillation with viscous-viscoelastic damping plots**

Figures 5, 7 represent the displacement–time graphs for case-1 and case-2; and figure 6 presents the velocity-time graph (for case-1) of the fractional continuously-variable order mass-spring damper models for the different values of damping ratio $\eta$ . The figure-5, 7 clearly brings out the difference in oscillatory character for the two different cases of continuously variable fractional derivative order in the system, for higher values of $\eta >> 0$ .The case-1 shows presence of abrupt change in rate of position a kink, whereas the case-2 has rather smooth decaying oscillation. It is noted here that, the initial position state i.e. $x(t)$ with increase of time tend to asymptotic with the equilibrium state that is $x=0$ . From the above figures, we observe the following:

i. When $\eta = 0.05$ or near to or equal zero, the system oscillates at its natural frequency $\omega_n$ and the system is called 'un-damped'. That means the oscillation will continue almost forever, like simple harmonic motion.

ii. When $\eta = 0.5$, the system oscillates at higher than the natural frequency and the system is called 'under-damped', like classical damped integer order oscillator. In this situation the oscillation gradually tends to zero. However relation of damped natural frequency say $\omega_d$ to $\omega_n$ in the case of continuously variable order damping case is to be developed; like we have in for integer order systems i.e. given by $\omega_d = \omega_n\sqrt{1-\eta^2}$ .

iii. When $\eta = 1$, the system oscillates quickly and it converges to zero as quickly as possible and the system is similar as called 'critical damped', like in integer order damped oscillators. In this situation the oscillation returns to equilibrium in the shortest period of time.

iv. When $\eta = \sqrt{\pi}$ , the system oscillates a littleas compared to critical damping and it converges to zero slowly and the system is similar as called 'over damped', like in integer order damped oscillators.

We point out here, the definitions regarding natural frequency, damped-frequency, under-damped oscillation, critically-damped oscillation, and over-damped oscillations in the continuously variable fractional order damped oscillator, that we have developed needs to be re-defined with respect to the variable fractional order $q$ of the system. But here we have drawn similarity with the classical integer order damped oscillator system.

## 7. Conclusion

In this article, we modelled the fractional continuously-variable order mass-spring damper systems for free oscillation with viscoelastic damping and forced oscillation with viscous-viscoelastic damping. The approach is new in the sense of changing of behaviour of guide continuously with the small change of time $\Delta t$ with respect to both viscoelastic and viscous-viscoelastic oscillator of order $q$. We also find the analytical solutions of the fractional continuously-variable order mass-spring damper systems by the successive iteration method. The graphical plots also have been presented for different values of damping ratio. From the graph we have given the conclusion for the nature of damping viz. undamped, underdamped, critical damping and over damping of the system, similar to those for integer order classical damped oscillator system, however, though these parameters in context of continuously variable order damping oscillator systems need to be developed. From this new method developed in this article we conclude that the proposed method is highly effective for finding the solution for the fractional dynamic model, where the fractional order of damping changes continuously. This development has immense potential in study of various physical dynamic systems.

## Acknowledgements

This research work was financially supported by BRNS, Department of Atomic Energy, Government of India vide Grant No. 2012/37P/54/BRNS/2382 for Research Project “Applications of Analytical Methods for the solutions of Generalized Fractional and Continuous order Differential Equations with the implementation in Computer Simulation”.

## References

[1] Miller K. S., Ross B., 1993, “An Introduction to the Fractional Calculus and Fractional Differential Equations”, John Wiley and Sons, New York.

[2] Podlubny I., 1999, “Fractional Differential Equations”, Academic Press, San Diego.

[3] Shen K. L., Soong T. T., 1995, “Modeling of Viscoelastic Dampers for Structural Applications”, J. Eng. Mech., 121, pp. 694–701.

[4] Koeller R. C., 1984, “Application of Fractional Calculus to the Theory of Viscoelasticity”, ASME J. Appl. Mech., 51, pp. 299–307.

[5] Enelund M., Ahler L. M., Runesson K., Josefson B. L., 1999, “ Formulation and integration of the standard linear viscoelastic solid with fractional order rate laws”, International Journal of solid and structures, 36, pp. 2417-2442.

[6] Ingman D., Shzdalnitsky J., 2001, “Iteration method for equation of viscoelastic motion with fractional differential operator of damping”, 190, pp. 5027-5036.

[7] Das S., 2011, “Functional Fractional Calculus”, Springer, New York.

[8] Rossikhin Y. A., Shitikova M. V., 1997, “Application of fractional operators to analysis of damped vibrations of viscoelastic single-mass systems”, Journal of Sound and vibration, 199 (4), pp. 567-686.

[9] Enelund M., Josefson B. L., 1997, “Time-domain finite element analysis of viscoelastic structures with fractional derivative constitutive relations”, American Institute of Aeronautics and Astronautics Journal, 35, pp. 1630-1637.

[10] Elshehawey E. F., Elbarbary E. M. F., Afifl N. A. S., El-Shahed M., 2001, “On the solution of theendolymph Equation Using Fractional Calculus”, Applied mathematics and computation, 124, pp. 337-341.

[11] Agrawal O. P., 2001, “Stochastic Analysis of dynamic system containing fractional derivatives”, Journal of Sound and vibration, 247(5), pp. 927-938.

[12] Miller K. S., 1993, “The Mittag-Leffler and Related Functions”, Integral Transforms and Special Functions, 1, pp. 41-49.

[13] Hong D. P., Kim Y. M., Wang J. W., 2006, “A New Approach for the Analysis Solution of Dynamic Systems Containing Fractional Derivative”, Journal of Mechanical Science and Technology, 20 (5), pp. 658-667.

[14] Gomez-Aguilar J. F., Rosales-Garcia J. J., Bernal-Alvarado J. J., Cordova-FragaT., Guzman-Cabrera R., 2012, “Fractional mechanical oscillators”, Revista Mexicana de Fisica, 58, pp. 348–352.

[15] Makris N., Constaninou M. C., 1991, “Fractional Derivative Maxwell Model for Viscous Dampers”, J. Struct. Eng., 117 (9), pp. 2708–2724.

[16] Choudhury M. D., Chandra S., Nag S., S Das, Tarafdar S., 2012, “Forced spreading and rheology of starch gel: Viscoelastic modeling with fractional calculus”, Colloids and Surfaces A: Physicochemical and Engineering Aspects, 407, pp. 64-70.

[17] Gaul, L., Klein P., Kemple S., 1989, “Impulse Response Function of an Oscillator with Fractional Derivative in Damping Description”, Mech. Res. Commun., 16 (5), pp. 297–305.

[18] Gaul L., Klein P., Kemple S., 1991, “Damping Description Involving Fractional Operators”, Mech. Syst. Signal Process, 5, pp. 8–88.

[19] Shokooh A., Suarez L., 1994, “On the Fractional Derivative Modeling of Damping Materials”, Technical Report presented for NASA Langley, College of Engineering, University of Puerto Rica, Mayaguez, PR.

[20] Bagley R. L., Torvik P. J., 1983, “Fractional Calculus- a Different approach to the analysis of viscoelastically damped structures”, AIAA Journal, 21, pp. 741-748.

[21] Suarez L., Shokooh A., Arroyo J., 1995, “Finite Element Analysis of Damping Materials Modeled by the Fractional Derivative Method”, Technical Report presented for NASA Langley, College of Engineering, University of Puerto Rico, Mayaguez, PR.

[22] Suarez L. E., Shokooh A., 1995, “Response of Systems With Damping Materials Modelled Using Fractional Calculus”, ASME J. Appl. Mech. Rev., 48 (11), pp. S118–S127.

[23] Saha Ray S., Bera R. K., 2005, “Analytical Solution of a Dynamic System Containing Fractional Derivative of Order One-Half by Adomian Decomposition Method”, Trans. ASME, 72, pp 290-295.

[24] Suarez L. E., Shokooh A., 1997, “An Eigenvector Expansion Method for the Solution of Motion Containing Fractional Derivatives”, ASME J. Appl. Mech., 64, pp. 629–635.

[25] Naber M., 2010, “Linear Fractionally Damped Oscillator”, International Journal of Differential Equations”, 2010, Article ID 197020.

[26] Stanislavsky A. A., 2004, “Fractional Oscillator”, Physical Review E, 70, 051103.

[27] Nutting P. G., 1921, “A New Generalized Law of Deformation”, Journal of the Franklin Institute, 191, pp. 679-685.

[28] Caputo M., 1969, “Elasticita e Dissipazione”, Zanichelli, Bologna.

[29] Slonimsky G. L., 1967, “Laws of Mechanical Relaxation Process in Polymers”, Journal of Polymer Science, 16, pp.1667-1672.

[30] Torvik P. J. Bagley R. L., 1984, “On the Appearance of the Fractional Derivative in the Behaviour of Real Materials”, Journal of Applied Mechanics, 51, pp. 294-298.